# Critical transverse compressive stresses of straight and bent CORC® wires with and without impregnation

X Xu[*], F Wan, S Cohan, V V Kashikhin

Fermi National Accelerator Laboratory, Batavia, IL 60510, U.S.A

[*]Author to whom any correspondence should be addressed. E-mail: xxu@fnal.gov

**Abstract**

CORC® wires are a promising superconductor for accelerator magnet applications. While their excellent uniaxial tensile properties have been well established, potential degradation under transverse compression remains a significant concern for accelerator magnets, in which transverse compression is a primary stress experienced by superconductors. To evaluate the critical transverse compressive stresses of ReBCO conductors, we developed an experimental system that enables testing of samples both with and without impregnation in liquid nitrogen. Furthermore, because bending strain induced during coil winding may influence the critical compressive response of CORC® wires, the apparatus was also modified to allow testing under the bending condition. In this study tests were conducted for five configurations: (1) straight wires without impregnation, (2) bent wires without impregnation, (3) straight wires with Stycast 2850 FT impregnation, (4) bent wires with Stycast 2850 FT impregnation, and (5) bent wires impregnated with paraffin wax. The transverse pressures corresponding to 3% and 5% reductions in the critical current are reported. The effects of wire bending and impregnation on the critical transverse pressure are analyzed, and the implications for transverse stress levels in accelerator magnet conductors are discussed.

**Keywords:** CORC® wire, ReBCO cables, transverse compression, irreversible degradation.

**1. Introduction**

Rare-earth barium copper oxide (ReBCO) high-temperature superconductors offer significant potential for a wide range of applications. Compared to the relatively mature low-temperature superconductors (LTS) such as Nb-Ti and $Nb_3Sn$, ReBCO conductors can operate at both higher fields and higher temperatures. For example, if the AC losses of present ReBCO conductors can be sufficiently reduced, accelerator magnets based on such ReBCO conductors operated at 20 K or above could achieve lower power consumption than $Nb_3Sn$ magnets operated at 1.9-4.5 K owing to the higher Carnot efficiency at elevated temperatures. Presently extensive efforts are underway within the community to develop ReBCO dipole magnets (e.g., [1-7]). Compared with other applications, accelerator magnets impose particularly demanding requirements on the conductors: e.g., high engineering current density ($J_e$), good electromechanical properties, low magnetization, and the ability to withstand small bending radii for most magnet designs. To reduce coupling-current magnetization, transposed cables are desired, while twisted cables are also employed when full transposition is difficult to realize. One of the most promising ReBCO cables for accelerator magnets is the conductor-on-round-core (CORC®) design, in which multiple layers of tapes are helically wound around a core, providing a twist of tapes in each layer [8]. By using tapes with thin substrates (30 μm or less), which allow small critical bending radii, cores with diameters below 3 mm can be used [9]. This enables CORC® wires to be bent to small diameters (below 30 mm) with minimal degradation of the critical current ($I_c$), and their round geometry results in direction-independent bending performance. The feasibility of CORC® wires for constructing dipole magnets has been clearly demonstrated by the canted-cosine-theta (CCT) magnets, with a bore field of 6 T achieved in the recent "C3" magnet [10,11].

It is well known that the achievable fields in superconducting magnets are limited not only by the current-carrying capability of the superconductors, but also by their mechanical properties, especially for high-field magnets in which Lorentz forces are substantial. CORC® wires and cables have been shown to possess excellent uni-axial tensile properties [12]. However, for accelerator magnets the dominant stress on the superconductors is transverse compression. At present dipole magnets developed using CORC® wires operate well below 10 T, in which the transverse compressive stress on the conductors is still low and not yet a big concern. However, for future dipole magnets with higher target operational field (15-20 T), whether in ReBCO-only or ReBCO/LTS hybrid configurations, it remains an open question whether the critical transverse pressure that the CORC® wires can withstand will become a limiting factor for the achievable field. This makes evaluation of the critical transverse compressive stress of CORC® wires a critical task. To apply a sufficiently high transverse compressive load to the conductor for this test, the most realistic method is by applying external pressure via anvils. Such tests were carried out by van der Laan et al. at 76 K for straight, non-impregnated CORC® wires and cables [13]. The results indicate that the critical transverse pressure of CORC® wires is significantly lower than that of ReBCO tapes [14]. An important reason for this is the presence of gaps between tapes within each layer in a cable, which are required to accommodate tape sliding during cable bending, but these gaps lead to stress concentrations in the tapes of adjacent layers when the cable is under high transverse pressure [13].

On the other hand, future accelerator magnets made from ReBCO conductors will likely require impregnation, as ReBCO is a brittle material and the conductor positions must be fixed to ensure adequate field quality. Therefore, it is essential to evaluate the transverse pressure limits of impregnated CORC® wires. In addition, bending strains generated when CORC® wires or cables are wound into coils may further affect the conductors' tolerance to applied stresses. Shen *et al.* [15] reported that the critical tensile strength of a bent Bi-2223 tape was only ~40% of that of a straight tape, indicating a strong coupling between bending strain and the tolerance to mechanical loads. Therefore, the critical transverse pressure of CORC® wires under the bending condition must also be assessed. To enable these investigations, an experimental system has been developed to test CORC® wires under transverse compression, with or without impregnation, in both straight and bent configurations. The corresponding experimental results are presented and discussed in this paper.

**2. Experimental**

*2.1. Samples*

Two CORC® wires fabricated by Advanced Conductor Technologies (ACT) LLC in 2020 were used for this study, with ID numbers of 191220-a and 191220-b (hereafter referred to as "a" and "b", respectively). Each wire consists of 29 ReBCO tapes, each 2 mm wide, tightly wound around a 2.55-mm-diameter OFHC copper core and arranged in 12 layers. The tapes were produced by SuperPower based on the AP formular, each with a 30-µm-thick Hastelloy substrate and 5-µm-thick electroplated copper on each side. The composite wires were enclosed in thin heat-shrink tubing. The resulting CORC® wires have an overall diameter of approximately 3.7 mm. The two CORC® wires were fabricated at the same time – the only difference between them is that different batches of SuperPower tapes were used.

*2.2. Measurements*

The experimental system developed in this work for transverse compression testing of ReBCO cables in liquid nitrogen is based on a Transverse Pressure Instrument (TPI) probe at Fermilab [16]. A picture of the testing setup is shown in Figure 1(a). The hydraulic actuator (the orange component in Figure 1a) is capable of generating forces of up to 20 US tons (about 178 kN). The hydraulic force is supplied and controlled by a hydraulic pump. A relief valve regulates the oil pressure in the system, which is measured using a calibrated pressure transducer. The applied force is calculated from the pressure reading and the known piston area of the hydraulic cylinder. The calculated forces were cross-checked against pressure distributions measured using Fuji pressure-sensitive films, showing good agreement. Figure 1(b) shows the lower end of the TPI probe. Two rods are housed inside the main probe cylinder; both the rods and the main cylinder are fabricated from Inconel 718 alloy. The hydraulic force is transmitted by pulling the two rods upward, and they are free to slide within the main cylinder. The lower ends of the rods are threaded to accommodate nuts. The sample holder is positioned between the bottom surface of the main cylinder and the nuts, so that the hydraulic force applies a compressive load to the holder. As shown in Figure 1(b), the holder consists of three parts, labeled "A", "B", "C", each containing two through-holes for the two Inconel rods. The diameters are 80 mm for Parts A and B, and 82 mm for Part C, respectively. Part A serves as the press die, whereas Parts B and C form the lower section of the pressing holder. Part B contains a groove that accommodates the samples. Parts B and C are interfaced via a conical surface, which allows part B to slide within the "bowl" of part C if there is any misalignment between the holder and the probe during loading. This design helps to keep the applied compression force aligned with the probe axis and normal to the groove. The groove is 4.5 mm wide, 10.1 mm deep, and 80 mm long. The press blade is 4.3 mm wide, 7.6 mm high, and 45 mm long. The blade length was longer than the twist pitch of the CORC® wires to ensure that the applied compression is uniform over at least one full twist pitch.

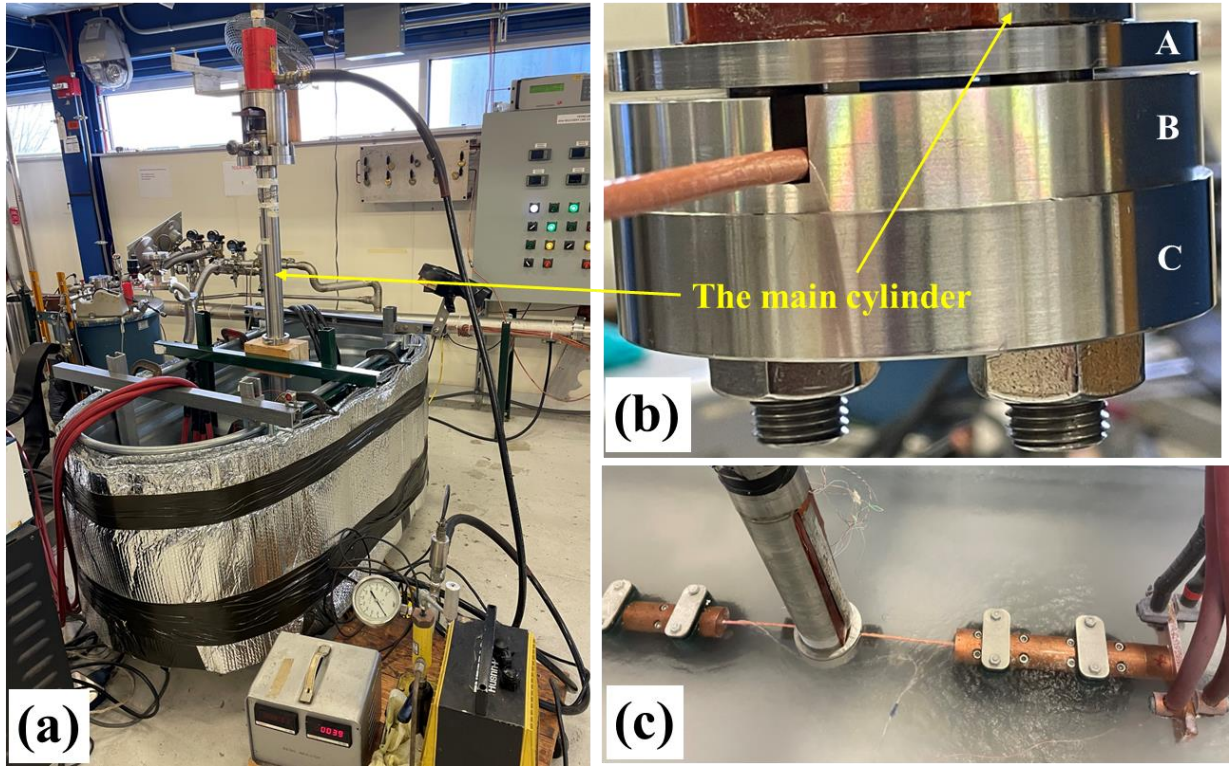

Figure 1. Pictures of the apparatus for the transverse pressure testing.

A photograph of a sample during testing is shown in Figure 1(c). Each test requires a CORC® wire about 60 cm in length. Each end of the wire was first prepared by exposing all 12 tape layers over a length of 20 cm (i.e., by trimming back the outer-layer tapes). Each prepared end, together with two copper wires serving as voltage taps, was inserted into a flute-type [11] OFHC copper tube (4.6 mm inner diameter, 6.4 mm outer diameter). The copper tubes were subsequently filled with 99.99% indium at 160ºC, thereby soldering the wire ends to the copper tubes. Because the voltage taps were integrated into the Cu terminals, the tests in this work measured the voltages between the two terminals (i.e., across the entire active sample length – about 20 cm). After surface cleaning, the copper tubes were clamped into copper adapters using bolts, as shown in Figure 1(c). The wire, together with the two adapters, was then mounted onto a G-10 plate: the wire was placed in the groove of the holder that was positioned on a platform attached to the G-10 plate in a configuration that ensured the wire remained straight and free of bending, while the adapters were fixed to the G-10 plate. The G-10 plate assembly was subsequently transferred into a bathtub. Copper plates connected to the current leads were bolted onto the sides of the adapters for current injection. The TPI probe was then positioned such that the two Inconel rods passed through the holes in the holder, after which nuts were installed on the rod ends. Then liquid nitrogen was filled into the bathtub prior to the testing. For each sample a *V-I* curve was first measured without applied pressure to determine its initial $I_c$ value, $I_c(0)$. The transverse pressure was then increased in steps and a *V-I* curve was recorded at each pressure. For most tests, a pressure-release step was included to assess the reversibility of $I_c$ reduction: once a decrease in $I_c$ was observed, the applied pressure was fully released, and the *V-I* curve was measured at zero pressure to determine whether the $I_c$ returned to $I_c(0)$. Subsequently, the pressure was increased again until $I_c$ dropped significantly.

To test impregnated wires, the groove of the holder was filled with impregnation materials; in this work Stycast 2850 FT was used. A key challenge is that most impregnation materials shrink during curing or solidification, producing a concave top surface, which would lead to a non-uniform distribution in the applied pressure. To ensure that the top surface of the Stycast after curing is flat in the region to be pressed, a T-shaped part was inserted into the groove and held in position during curing so that its flat bottom surface defined the top surface of the cured Stycast. The vertical portion of the T-shaped part (i.e., the section inserted into the groove) was 5.1 mm high, 50 mm long, and 4.4 mm wide, closely matching the groove width. Its length was chosen to be slightly longer than the press blade, ensuring a flat surface over the entire pressed region while leaving certain lengths of the groove uncovered, through which liquid impregnation materials could be added into the groove. The impregnation procedure was as follows. After small shim pieces were installed at the ends of the groove (which would center the wire laterally and lift its bottom 0.5 mm above the groove floor), the wire was placed in the groove, and the ends of the groove were sealed with putty. Then sufficient Stycast was poured into the groove to make sure that the liquid level was above the desired final height, after which the T-shaped part was inserted into the central region of the groove, and more Stycast was added into the groove so that the liquid level outside the T-shaped insert rose well above the bottom surface of the insert. After curing, the T-shaped insert was removed, leaving a flat Stycast surface in the central region of the groove. A picture of a holder after the Stycast curing is shown in Figure 2(a), and a schematic showing the approximate Stycast dimensions is given in Figures 2(b).

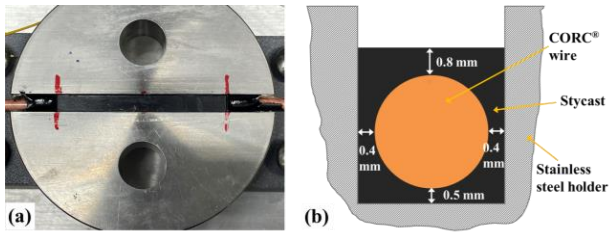

Figure 2. (a) A picture of a sample holder after the Stycast curing, and (b) a schematic showing the approximate dimensions of the cured Stycast.

A picture of the experimental setup to test samples in the bending configuration is shown in Figure 3(a). The wire ends were prepared, soldered into Cu tubes, and clamped into the adapters following the same procedure described above. The sample was then tightly bent around a former with a radius of 50 mm (not shown in Fig. 3). While maintaining this curvature, the two adapters were clamped onto a G-10 plate, after which the former was removed. The bent sample was subsequently placed into the holder, which was mounted on a platform attached to the G-10 plate. The groove used for the bending tests is 4.7 mm wide and 8 mm deep. The press blade is 4.5 mm wide, 5 mm high, and about 43 mm long. To test the bent samples with impregnation, the same impregnation procedure described above was used. A T-shaped insert with a 4.5 mm wide, 3 mm high, and 50 mm long vertical section was used to ensure that the impregnation material surface was flat. The curvature of the groove, the press blade, and the T-shaped insert was designed to match that of the bent wire. To study the influence of impregnation material properties, pure paraffin wax (i.e., without added fillers) with a Durometer 40D hardness was used in addition to Stycast 2850 FT. A picture of the holder with wax impregnation is shown in Figure 3(a). A schematic showing the approximate dimensions of the impregnation materials after curing is given in Figure 3(b).

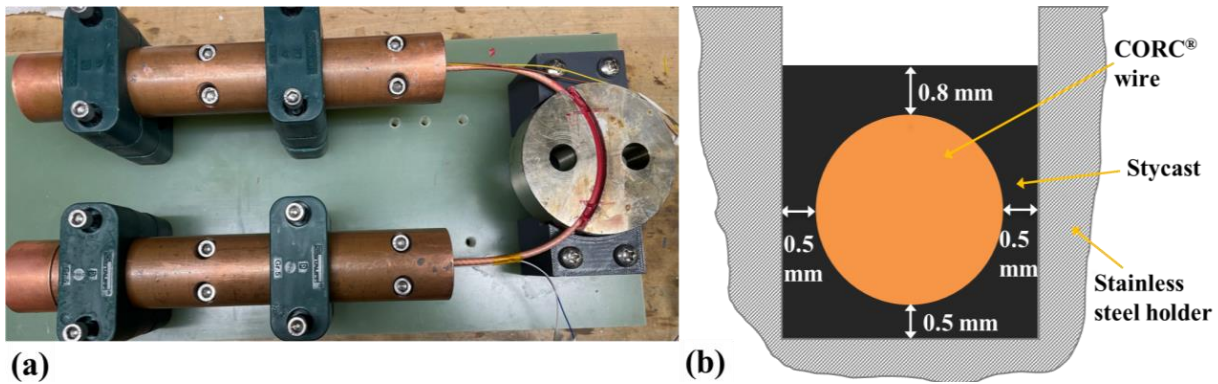

Figure 3. (a) A picture of the setup for testing a sample in the bending configuration with wax impregnation, (b) a schematic showing the approximate dimensions of the impregnation material after curing.

## 3. Results

*3.1. Testing of straight samples without impregnation*

The test results of straight CORC® wires without impregnation are shown in Figure 4. Two samples were tested: a "b" wire using a press die with sharp blade edges, and an "a" wire using a press die with rounded blade edges (~1 mm radius) to avoid stress concentration at the blade/wire interface. Figure 4(a) shows the *V-I* curves of the "b" sample measured under various transverse pressures. Because the contact area between a round wire and the press blade increases with applied compression, it is not possible to determine the compressive stress in the unit of MPa. Instead, here the load is normalized by the press blade length and expressed in the unit of kN/m. Figure 4 (b) shows their $I_c$ values as a function of the load. The "a" wire exhibits noticeably higher $I_c$ than the "b" wire. Although the ReBCO tapes used in these two wires were produced based on the same recipe and parameters, variation in $I_c$ among different batches of tapes leads to noticeably different $I_c$ values between the two CORC® wires. In fact, as will be shown later, even different segments from the same wire may exhibit slight variation in $I_c$. To compare the responses of the two wires to the transverse pressure, their $I_c$ values under pressures are normalized to their respective initial $I_c(0)$ values and are shown in Figure 4(c). It is seen that the two curves more or less overlap.

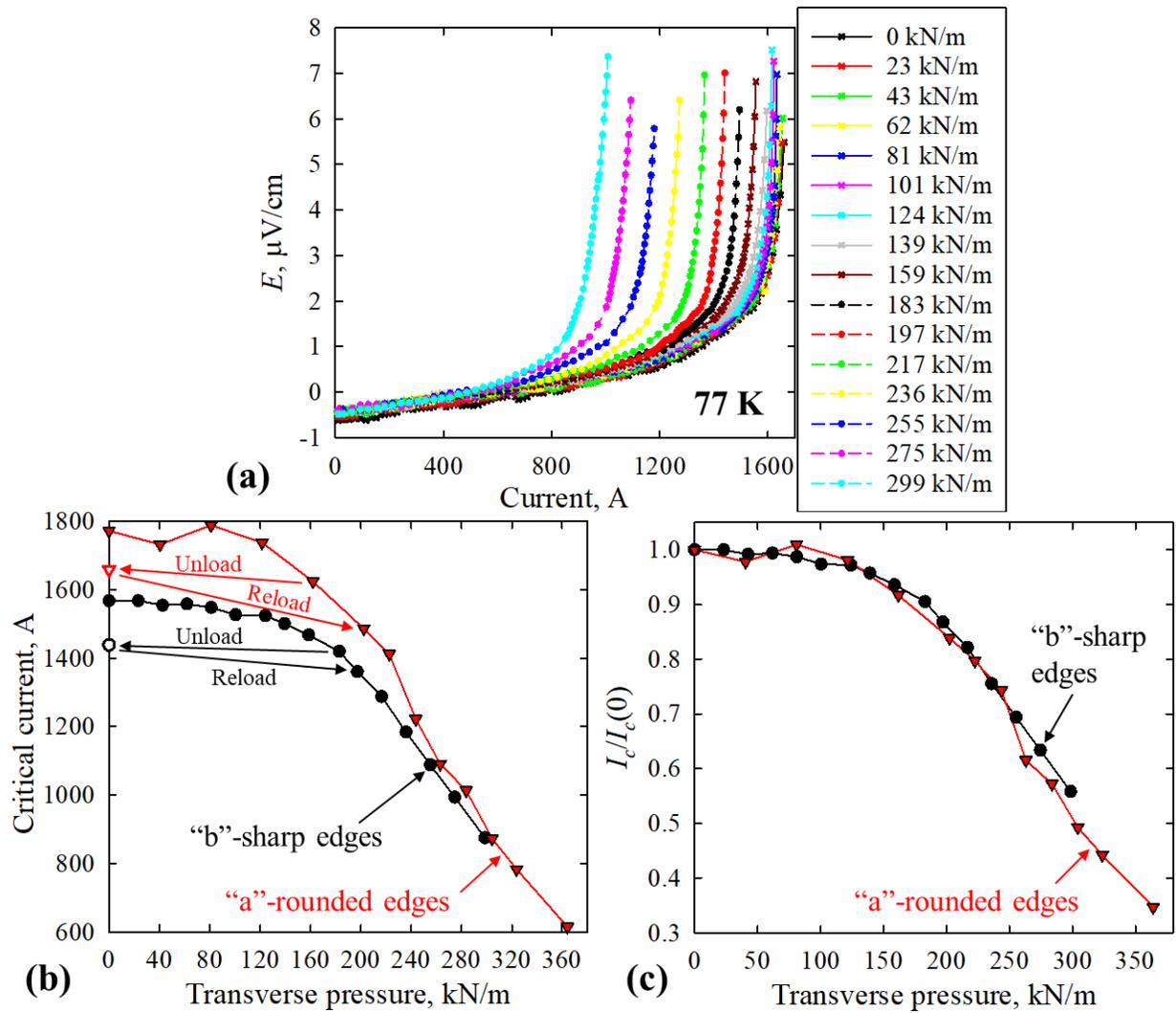

Figure 4. Test results for straight CORC® wires without impregnation: (a) *V-I* curves of the "b" wire under various transverse compression loads, (b) the $I_c$ values as a function of the compression loads, and (c) normalized $I_c$ as a function of the compression loads.

*3.2. Testing of bent wires without impregnation*

For the bending configuration (bending radius of 50 mm), a "b" wire and an "a" wire were tested using a press die with sharp blade edges, while an additional "a" wire was tested using a press die with rounded blade edges (~1 mm radius). The *V-I* curves of the "a" wire tested using the press with sharp blade edges are shown in Figure 5(a). The calculated $I_c$ values as a function of transverse load (in kN/m) for the three samples are shown in Figure 5(b). It is seen that the "b" wire experienced a significantly larger $I_c(0)$ drop after bending compared with the "a" wire, and that there was a slight $I_c(0)$ variation between the two segments of the "a" wire. The $I_c$ values under pressure are normalized to their $I_c(0)$ values and shown in Figure 5 (c).

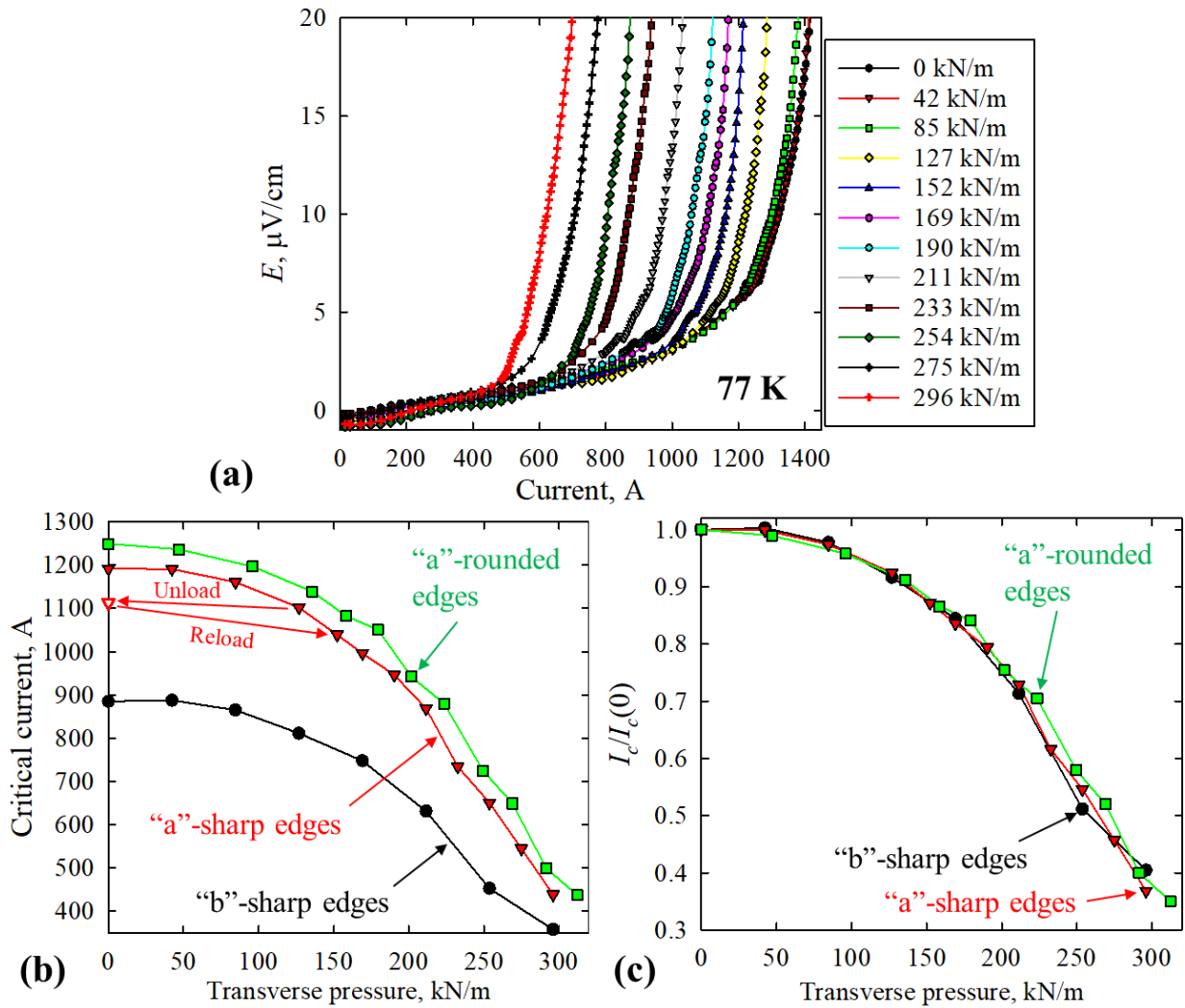

Figure 5. Testing results of bent wires without impregnation: (a) *V-I* curves of the "a" wire tested using the press with sharp blade edges, (b) the $I_c$ values as a function of the compression loads, and (c) normalized $I_c$ as a function of the compression loads.

*3.3. Testing of a straight wire with Stycast 2850 FT impregnation*

A straight "a" wire was tested with Stycast 2850 FT impregnation, and the results are shown in Figure 6. The *V-I* curves at the various pressures are shown in Figure 6(a). For the impregnated samples, because the pressed area is just equal to the area of the press blade, the applied compressive stress is calculated as the force divided by the blade area and is expressed in MPa. The calculated $I_c$ values as a function of the transverse compressive stress are shown in Figure 6(b).

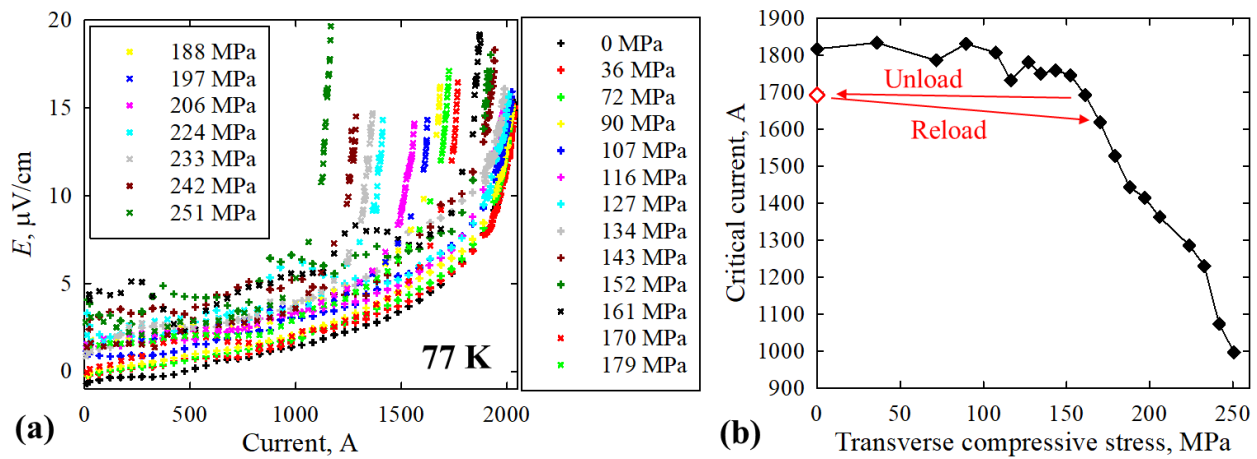

Figure 6. Test results for a straight "a" wire with Stycast impregnation: (a) *V-I* curves under various transverse pressures, and (b) the $I_c$ values as a function of the compressive stress.

*3.4. Testing of bent wires with impregnations*

An "a" wire was tested in the bending configuration with Stycast 2850 FT impregnation. The *V-I* curves at various transverse pressures are shown in Figure 7(a), and the calculated $I_c$ values as a function of the pressure are shown in Figure 7(b).

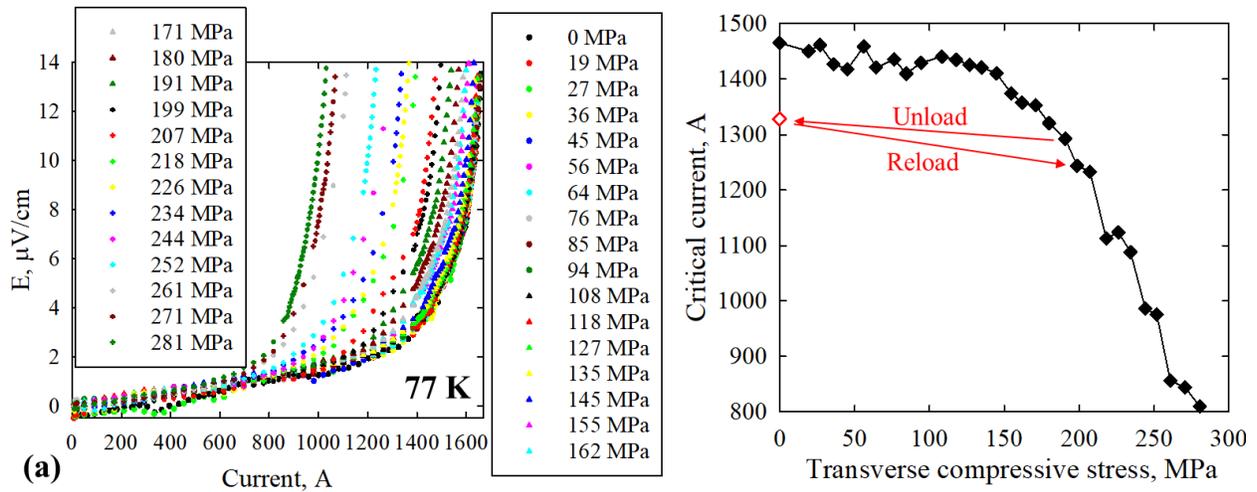

Figure 7. Test results of an "a" wire in the bending configuration with Stycast impregnation: (a) *V-I* curves under various transverse pressures, and (b) the $I_c$ values as a function of the compressive stress.

In addition, an "a" wire was tested in the bending configuration with paraffin wax impregnation. The *V-I* curves and the calculated $I_c$ values as a function of the applied pressure are shown in Figure 8 (a) and (b), respectively.

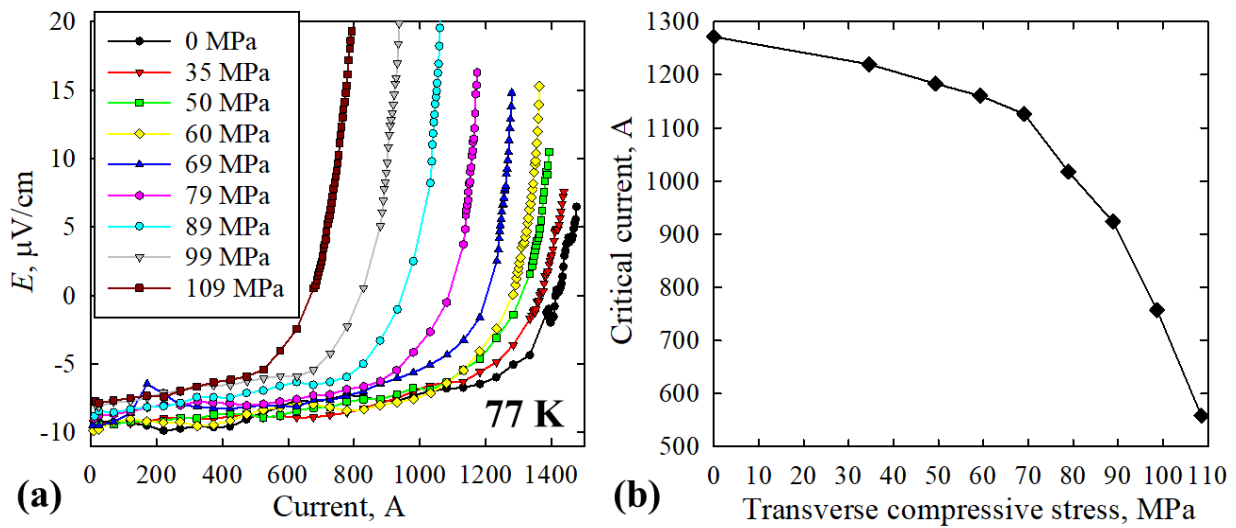

Figure 8. Test results for an "a" wire in the bending configuration with paraffin wax impregnation: (a) *V-I* curves and (b) the $I_c$ values as a function of transverse pressure.

## 4. Discussions

By comparing the $I_c(0)$ values of the straight and bent wires, it is seen that the "a" and "b" wires lost roughly 20-30% and 40% of their $I_c$s after bending, respectively. It should be noted that these two wires were produced by ACT in 2020 using the previous generation of lubricants, which degraded over time leading to an inferior bending performance. A number of FEM studies have shown that a low friction coefficient is critical for CORC® wires to achieve small critical bending radii (e.g., [17,18]). The CORC® wires produced in recent years have adopted new lubrication materials that are free of this issue, and thus exhibit significantly improved bending performance [19].

As shown in Figure 5(c), the three curves of normalized $I_c$ as a function of the transverse load overlap, indicating that the use of sharp or rounded blade edges has little influence on the critical transverse pressure. This suggests that the $I_c$ degradation was primarily governed by the transverse pressure within the pressed region, rather than by stress concentrations at the blade edges. Furthermore, although the two CORC® wires, "a" and "b", have noticeably different $I_c(0)$ values, their sensitivities to transverse pressure are similar, as is also evident from Figure 4(c).

The results of this study further indicate that there is no pressure regime in which the reduction in $I_c$ is reversible. As shown in Figs. 4–7, once a decrease of a few percent in $I_c$ occurred, the subsequent release of the applied pressure did not restore $I_c$ to its original value $I_c(0)$; instead, $I_c$ remained close to the reduced level observed immediately prior to unloading. This behavior indicates that irreversible degradation had already occurred. This behavior contrasts with that of $Nb_3Sn$ superconductors, in which a reversible $I_c$ reduction typically precedes irreversible degradation under transverse compressive loading [20]. Conversely, the response of CORC® wires is similar to that of Bi-2212 cables, for which $I_c$ degradation under transverse pressure has also been reported to be entirely irreversible [21].

A summary of the transverse pressures corresponding to 3% and 5% reductions in $I_c$ for the five test configurations is shown in Table 1. The 3% and 5% criteria are selected here because it was reported in [13] that for a straight, non-impregnated CORC® wire with 2.55-mm-diameter Cu core, a monotonic loading that caused an $I_c$ decrease of 5% or above would lead to further degradation under subsequent cyclic loadings of the same magnitude. In the present work the samples were subjected only to one-off loads. However, given that conductors in accelerator magnets need to experience many loading-unloading cycles, it is necessary to extend the transverse compression tests to include cyclic loading for the bent and impregnated conductors.

Table 1. Summary of transverse pressure loads corresponding to 3% and 5% decrease in $I_c$. Note that due to the limited number of data points, most values were estimated by interpolation.

| Samples | 3% | 5% |
| --- | --- | --- |
| Straight wires without impregnation, kN/m | 125 | 145 |
| Bent wires without impregnation, kN/m | 90 | 104 |
| Straight wires with Stycast 2850 FT impregnation, MPa | 140 | 156 |
| Bent wires with Stycast 2850 FT impregnation, MPa | 134 | 150 |
| Bent wires with paraffin wax impregnation, MPa | 25 | 40 |

Table 1 shows that both wire bending and impregnation have noticeable influences on the critical transverse compressive tolerance of CORC® wires. While finite element modeling (FEM) is required for a quantitative assessment of these effects, a simplified qualitative analysis is presented below to provide physical insight into the underlying mechanisms.

Regarding the influence of wire bending, Table 1 shows that the critical transverse pressures of the bent wires are somewhat lower than those of the straight wires, but the effect of bending observed here is significantly less pronounced than that reported in [15] for the tensile test of a Bi-2223 tape. This is probably because in the study [15] both the bending and the tensile loading generated longitudinal tensile strains in the Bi-2223 tape, so these two strain components directly superimposed. Because the tape can tolerate only a limited longitudinal tensile strain, bending consumed part of this strain margin, resulting in a reduced critical tensile stress. To understand the influence of bending on the critical transverse pressure of CORC® wires, it is necessary to examine the strain components induced by bending and by transverse compression, and to assess whether they superimpose. Bending of a CORC® wire generates tensile and compressive strains within the tapes, primarily in the tape plane [22]. In contrast, transverse pressure predominantly induces compressive strains normal to the tape plane. These out-of-plane compressive strains do not directly superimpose on the in-plane strains induced by wire bending. Instead, through the Poisson effect, the out-of-plane compressive strains generate in-plane tensile strains, which add to the in-plane tensile strain induced by wire bending. Therefore, bending of a CORC® wire can reduce its tolerance to transverse compression, but because the in-plane tensile strain generated by transverse pressure arises as a secondary effect of the Poisson response and is relatively small, the influence of bending here is less severe than that for the Bi-2223 tape shown in [15]. It should also be noted that wire bending may affect the transverse compression load limits through additional mechanisms. For example, bending increases the gap spacing between tapes within the same layer on the outward side (i.e., the side farther from the bending center). As shown in [13], larger gap spacing can lead to lower transverse compression load limits due to stress concentration. Furthermore, any damage incurred during bending (e.g., tape wrinkling if inter-tape sliding is hindered by a high friction coefficient [22]) may further degrade performance under transverse compression.

Regarding the influence of impregnation, Table 1 lists transverse pressure limits in units of kN/m for non-impregnated wires and in MPa for impregnated wires; therefore, the values cannot be directly compared. If the values for the impregnated wires are converted from MPa to kN/m, the critical transverse loads for the bent, Stycast-impregnated, wire (600–670 kN/m) are over six times higher than those for the bent, non-impregnated, wire, whereas those for the wax-impregnated wire (110-180 kN/m) are only 20-70% higher. This demonstrates that a stiff impregnation material can significantly enhance resilience to transverse pressure, while the influence of a soft impregnation material (such as pure paraffin wax) is only marginal. This behavior can be attributed to two primary factors. First, when an impregnated wire is subjected to transverse pressure, its entire outer surface participates in load bearing, and according to an FEM study for impregnated, round, $Nb_3Sn$ wires under transverse compression, the generated pressure is approximately normal to the wire surface in all directions [23]. Second, the applied load is not carried solely by the conductor; a portion is transferred to and supported by the impregnation material. According to the rule of mixtures, the effective modulus of an impregnated composite depends on the volume fractions of the conductor and the impregnation material as well as their respective elastic moduli. This effective modulus, in turn, governs the overall strain of the composite under an applied transverse pressure.

To understand how different impregnation materials influence the effective elastic modulus of impregnated CORC® wires, stress–strain behavior under transverse compression was measured for several samples using the straight-sample holder shown in Fig. 1(b). The reduction in sample thickness with increasing load was determined by measuring the decrease in the gap between Parts A and B of the holder using a thickness gauge (with a resolution of 0.01 mm), with the deformation of the stainless-steel press blade subtracted. The measurements were performed at room temperature. Four sample configurations were investigated: (1) a CORC® wire impregnated with Stycast 2850 FT, (2) Stycast 2850 FT alone (i.e., Stycast poured into the groove without a wire, with all other conditions identical), (3) a CORC® wire impregnated with paraffin wax, and (4) paraffin wax alone. However, significant creep of the wax was observed under high compressive stress at room temperature, preventing meaningful stress–strain characterization for cases (3) and (4). The results for the first two configurations are shown in Figure 9. Linear fits were applied to the stress–strain data to extract elastic moduli. As reported in previous studies [24,25], the transverse stress–strain response of a CORC® wire typically exhibits an initial low-modulus region associated with the progressive closure of gaps between tape layers. Accordingly, the low-stress region was excluded from the linear fit for the impregnated CORC® wire in Figure 9. The obtained elastic modulus for the Stycast 2850 FT is 8.6 GPa, somewhat higher than those reported in literature [26]. The obtained effective modulus for the Stycast-CORC® composite is 6.0 GPa, which is much higher than those reported in [25].

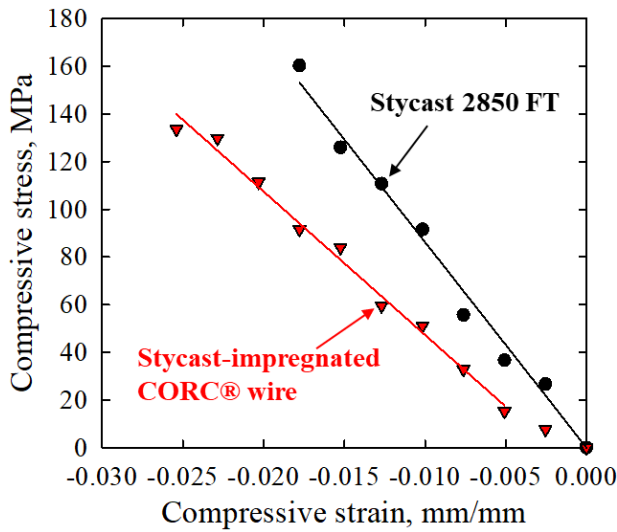

Figure 9. Transverse compressive stress-strain test results for the Stycast 2850 FT and a Stycast-impregnated CORC® wire measured at room temperature.

The reason why the moduli obtained in this work are higher is probably that the cured Stycast was bonded to the stainless-steel holder, which to some extent constrained Stycast deformation (in comparison, free-standing Stycast-wire composites were tested in [25]). It should be noted that in impregnated coils the impregnation material is also bonded to the structural materials. The composite modulus measured in this work is lower than that of Stycast alone, indicating – based on the rule of mixtures – that the modulus of the CORC® wire itself in the impregnated configuration is substantially lower than that of Stycast. Furthermore, at cryogenic temperatures the modulus contrast between Stycast and CORC® wires becomes even more pronounced. As temperature decreases from room temperature to 77 K, the elastic modulus of Stycast can increase by a factor of three [27], whereas the moduli of most metals and alloys increase only weakly, typically by 10% or less [28] – a similar behavior is expected for CORC® wires. Compared with Stycast, paraffin wax is significantly softer, with an elastic modulus of approximately 0.1-0.2 GPa at room temperature and 2-4 GPa at 77 K [29]. This much lower modulus leads to a lower effective modulus of the wax–conductor composite and, consequently, a higher strain for a given transverse compressive stress. In addition, paraffin wax becomes brittle at 77 K – we observed many cracks in the wax after the compression tests. Given the large difference between the elastic modulus of the impregnation material and that of the CORC® wire, the groove geometry is expected to influence the transverse compression tolerance of impregnated CORC® wires. For example, in the case of Stycast impregnation, a wider groove increases the volume fraction of Stycast and thus the effective composite modulus. This higher modulus reduces the compressive strain under a given applied stress and is therefore expected to mitigate conductor degradation.

On the other hand, although the present results show that impregnating CORC® wires with a stiff material such as Stycast can significantly enhance their tolerance to transverse pressure, whether this improvement is beneficial in practice depends in fact on the magnet design. During magnet operation the Lorentz force per unit length acting on the conductor is: $F$ [kN/m] = $B$ [T]*$I$ [kA]. However, the actual stress experienced by the conductors can be substantially higher than this value, depending on the coil architecture and impregnation option.

- *Coils without stress management (e.g., standard cosine-theta or block-type coils).* In such coils, the Lorentz forces on individual turns accumulate within the winding pack. For example, the accumulation of azimuthal stress leads to the highest stresses in the conductors near the coil midplane, which can be significantly greater than the Lorentz force experienced by any individual turn of the conductors in the coil. Apart from the azimuthal stress, radial stress can also accumulate within a coil. Additional contributions from preload and differential thermal contraction may further increase the overall stress. To enable such coils to achieve high fields (e.g., > 10-12 T), impregnation of the conductors with a stiff material seems unavoidable in order to manage the high stress. In this case the critical transverse pressures measured for the impregnated wires shown in Table 1 (e.g., 134-150 MPa for the bent wires) represent the limits that the conductors can withstand and can be used for estimating the maximum achievable field.

- *Stress-management coils without impregnation.* In stress-management coils such as canted-cosine-theta (CCT) or conductor-in-molded-barrel (COMB) [3] designs, structural materials such as the ribs and spars intercept the Lorentz forces that each turn of the conductors experiences. These forces accumulate in the structural components, with the highest azimuthal stresses occurring in the spars near the coil midplane. If the conductors are not impregnated, because the grooves in the structural materials are typically larger than the conductors, the stresses in the structural materials cannot be transferred to the conductors. In this case, the load on a conductor is just equal to the Lorentz force $F = B*I$ acting on it. Accordingly, the critical transverse load limits measured for non-impregnated wires shown in Table 1 (around 100 kN/m for the bent wires) can be used for estimating the maximum achievable field.

- *Stress-management coils with impregnation*. If the grooves in the structural materials are filled with an impregnation material, the conductors become mechanically coupled to the surrounding structure. As a result, stresses in the structural components can be transferred to the conductors, causing the highest azimuthal stress in the conductors close to the midplane. In this case, the loads on the conductors can be much higher than the Lorentz force $F = B*I$ acting on them. Based on the rule-of-mixture, it can be calculated that the amount of load transferred from the structural materials to the conductors depends on the stiffness of the impregnation material: the stiffer the impregnation material, the greater the load transfer. In fact, the non-impregnated case discussed above can be viewed as a special case in which the impregnation material is "infinitely soft", resulting in negligible load transfer. Therefore, we see two conflicting factors determining the conductor stress in stress-management coils. On one hand, a stiffer impregnation material increases load transfer from the structural materials to the conductors, leading to higher stresses and strains. On the other hand, for a given Lorentz force $F = B*I$ acting on the conductor, a stiffer impregnation material increases the effective composite modulus and thereby reduces the conductor strain. In practice, the stress-strain state of the conductors in a coil is further influenced by factors such as differential thermal contraction, preload conditions, and assembly tolerances. Therefore, detailed magnet-specific mechanical design and stress analysis are required to determine the optimal impregnation strategy for a given magnet configuration.

Finally, improving the tolerance of CORC® wires to transverse pressure is important for reaching higher field for accelerator magnets. Based on results from previous studies (e.g., Ref. [13]) and the present work, the critical transverse pressure of a CORC® wire is believed to be governed primarily by three factors.

(1) *The ReBCO tapes used to fabricate the CORC® wire*. It is understandable that there is a strong correlation between the critical transverse compressive stress of a CORC® wire and the constituent tapes. For example, increasing the Cu stabilizer thickness in ReBCO tapes, which has been shown to increase their sensitivity to transverse pressure [14], also reduces the critical transverse pressure of the corresponding CORC® wire [13].

(2) *The wire design*. As shown in [13], the critical transverse pressure of a CORC® wire can be improved by reducing the gap spacing between tapes within each layer or by increasing the core size. However, both design modifications adversely affect the bending performance of the wire. In addition, there may be other parameters such as the tape winding angle and friction coefficient that may influence the critical transverse pressures. Therefore, the CORC® wire design must be optimized to achieve both good bending performance and high transverse pressure tolerance.

(3) *The conductor condition and operating configuration*. The mechanical state of the conductor, including factors such as bending radius and impregnation condition, also plays a critical role. Optimizing coil design parameters and impregnation strategies is therefore essential to enhance transverse pressure tolerance under realistic magnet operating conditions.

## 5. Conclusions

In this work, an experimental system was developed to measure the critical transverse compressive tolerance of CORC® wires. Five configurations were investigated: (1) straight wires without impregnation, (2) bent wires without impregnation, (3) straight wires impregnated with Stycast 2850 FT, (4) bent wires impregnated with Stycast 2850 FT, and (5) bent wires impregnated with paraffin wax. The results show that wire bending leads to a modest reduction in the critical transverse pressure, but the effect is not severe. Non-impregnated wires exhibit relatively low transverse pressure limits (on the order of 100 kN/m for the bent wires), whereas impregnation with a stiff material such as Stycast 2850 FT significantly increases the critical transverse pressure (to 134-150 MPa for the bent wires). In contrast, impregnation with a soft material such as paraffin wax provides only a marginal improvement. The analysis further indicates that the optimal choice of impregnation depends strongly on the magnet design, including whether stress-management features are employed. The experimental results presented in this work provide valuable input for magnet mechanical design, enabling informed selection of conductor impregnation strategies and estimation of the maximum achievable field constrained by the transverse pressure limits of CORC® wires.

**Acknowledgements**

This work was produced by Fermi Forward Discovery Group, LLC under Contract No. 89243024CSC000002 with the U.S. Department of Energy, Office of Science, Office of High Energy Physics. Publisher acknowledges the U.S. Government license to provide public access under the DOE Public Access Plan. The authors would like to thank Danko van der Laan at ACT

LLC for valuable discussions. The authors also thank Vito Lombardo at Fermilab for providing the CORC® wires used in this work.**References**

[1]. Durante M, Borgnolutti F, Bouziat D, Fazilleau P, Gheller J-M, Molinié F and Antoni P D 2018 Realization and first test results of the EuCARD 5.4-T REBCO dipole magnet *IEEE Trans. Appl. Supercond.* **28** 4203805

[2]. Wang X, Caspi S, Dietderich D R, Ghiorso W B, Gourlay S A, Higley H C, Lin A, Prestemon S O, van der Laan D and Weiss J D 2018 A viable dipole magnet concept with REBCO CORC® wires and further development needs for high-field magnet applications *Supercond. Sci. Technol.* **31** 045007

[3]. Kashikhin V V et al. 2024 Accelerator magnet development based on COMB technology with STAR wires *IOP Conf. Ser.: Mater. Sci. Eng.* **1301** 012153

[4]. Lecrevisse T, Benoist E, Blondelle A, Caunes A, Durochat M, Genot C, Lenoir G, Maloeuvre B, Ballarino A and Baskys A 2026 Metal as Insulation REBCO Racetracks Coils: Development, Fabrication, and Cryogenic Testing at CEA Paris-Saclay *IEEE Trans. Appl. Supercond.* **36** 4001405

[5]. Baskys A, Ballarino A, Barth C, Fiscarelli L, Gal N, Mangiarotti F, Mazet J, Perini D and Saba A 2026 Development and Testing of the First REBCO Racetrack Model Coils at CERN *IEEE Trans. Appl. Supercond.* **36** 4003306

[6]. Araujo D M, Auchmann B, Brem A, Daly M, Michlmayr T, Muller C, Rodrigues H G, Sotnikov D and Milanese A (2024) Subscale stress-managed common coil design *IEEE Trans. Appl. Supercond.* **34** 4003905

[7]. Q Xu 2025 Overview of HTS conductor and magnet development in China https://indico.cern.ch/event/1455509/timetable/#20251024.detailed

[8]. van der Laan D C 2009 YBa2Cu3O7−δ coated conductor cabling for low ac-loss and high-field magnet applications *Supercond. Sci. Technol.* **22** 065013

[9]. Weiss J D, Mulder T, ten Kate H J and van der Laan D C 2017 Introduction of CORC® wires: highly flexible, round high temperature superconducting wires for magnet and power transmission applications *Supercond. Sci. Technol.* **30** 014002

[10]. Wang X et al. 2020 Development and performance of a 2.9 Tesla dipole magnet using high-temperature superconducting CORC® wires *Supercond. Sci. Technol.* **34** 015012

[11]. Yan Y, Wang X et al., 2025 Fabrication and test results of a canted cosθ dipole magnet using high-temperature superconducting CORC® wires *International Conference on Magnet Technology* Wed-Af-Or1-01
25